# Francesco Fontana and his *astronomical* Telescope

Paolo Molaro
INAF-OATs
Via G.B. Tiepolo 11, I 34143, Trieste, Italy

## Abstract

In the late 1620s the Neapolitan telescope-maker Francesco Fontana was the first to observe the sky using a telescope with two convex lenses, which he had manufactured himself. Fontana succeeded in drawing the most accurate maps of the Moon's surface of his time , which were to become popular through a number of publications spread all over Europe but without acknowledging the author. At the end of 1645, in a state of declining health and pressed by the need to defend his authorship, Fontana carried out an intense observational campaign, whose results he hurriedly collected in his *Novae Coelestium Terrestriumque rerum Observationis* (1646), the only book he left to posterity. Fontana observed the Moon's main craters, as the crater Tycho which he named *Fons Major*, their radial patterns and the change in their positions due to the Moon's motions. He observed the gibbosity of Mars at quadrature and, together with the Jesuit Giovanni Battista Zupus, he described the phases of Mercury. Fontana observed the two - and occasionally three - major bands of Jupiter, and inferred the rotation movement of the major planets Mars, Jupiter and Saturn, arguing that they could not be attached to an Aristotelian sky. He came close to revealing the ring structure of Saturn. He also suggested the presence of additional moons around Jupiter, Venus and Saturn, which prompted a debate that lasted more than a hundred years. In several places of his book Fontana repeatedly claimed to have conceived the first positive eyepiece in 1608, providing a declaration by Zupus to have used his telescope since 1614. This declaration is still the oldest record mentioning such a device. We finally suggest that the telescopes depicted in the two paintings *Allegory of Sight* and *Allegory of Sight and Smell* by J. Brueghel the Elder belonging to Albert VII might have been made by Fontana, and that he might have inspired the *Allegory of Sight* by Jusepe Ribera (c. 1616).

## 1. Introduction

    The scarce information we have on Francesco Fontana is given by his contemporary Lorenzo Crasso, who in 1666 dedicated a book to the outstanding people of his time and counted Fontana among them. From Crasso's short biographical notes we learned that Fontana was born in Naples sometime between 1580 and 1590 and that at the age of 20 graduated in Theology and Law obtaining his Doctorate at the University of Naples Federico II. However, he never practiced in that profession and, following a vocation he had shown ever since his childhood, he self-taught mathematical sciences and devoted himself to grinding lenses. Crasso reported that Fontana used to say he preferred the truth of science to that of the Forum [1]. At the death of Giovan Battista Della Porta, considered by Fontana as the inventor of the telescope, he made several but all unsuccessful attempts to obtain Della Porta's instruments. In Naples he was close to Camillo Gloriosi, correspondent and, in 1610, successor of Galileo at Padua University, and with the Lyncean Fabio Colonna who commissioned him microscopic observations in 1625. Nevertheless, Fontana was also close to the Neapolitan Jesuits, who were frequently opposed to the Lynceans, in particular with fathers Girolamo Sersale, Giovan Batista Zupi and Giovan Giacomo Staserio.
    Fontana was a fine craftsman and never needed to do something else for a living. His telescopes reached the courts interested in scientific and military developments all over Europe. The quality of his lenses was so

high that in 1638 Fulgenzio Micanzio, in a letter to Galileo, wrote: *'Continual working on and construction of telescopes is said to have reached such unusual qualities that in matters of the heavens he is a miracle'* [2]. To advertise his telescopes, Fontana used to send maps of the Moon and news of other discoveries he had made by observing the sky from the roof of his house in Naples: *having manufactured for himself two of enormous length and fitted on a wooden support on the top of his house, with which observing constantly the planets*, formed the Book entitled Novae Observationes caelestium terrestriumque rerum, which he gave to light in 1646 (Crasso) [3].

The *'Novae Coelestium, Terrestriumque rerum observationes, et fortasse hactenus non vulgate a Francisco Fontana specillis a se inventis, et ad summam perfectionem perductis, editae"* (1646) is the only work he published, though Crasso mentioned a treaty, "*On Fortifications*", which has never been found. In 1656 Fontana died from plague along with all his numerous family, and all his mastery was lost.

The work of Fontana received little attention, if not open opposition, of scholars and scientific circles of all times, save for rare exceptions such as Colangelo (1834). His numerous detractors generally emphasized the superficiality, if not the incorrectness, of his observations and the lack of any optical theory for the functioning of the telescope. His claim to have constructed a telescope with two convex lenses in 1608 was generally considered as unreliable. On 25 May 1647 Torricelli wrote to Vincenzo Renieri: *'I have the book of stupidities observed, or rather dreamed up, by Fontana in the heavens'*. In his Almagestum Novum (1651), the Jesuit Giovanni Battista Riccioli while acknowledging the quality of the instruments constructed by Fontana, rejected most of the '*novelties'* Fontana had observed. The recent translation from Latin made by Beaumont and Fay (2001) allows us to make an accurate study of Fontana's writings which return a rather different, but probably more realistic, image of the scientist and of his work. The translation has been distributed privately and I here used the copy of the library of the Observatoire de Paris.

## 2. The Novae Coelestium observationes

The '*Novae Coelestium, Terrestriumque rerum observationes, et fortasse hactenus non vulgate a Francisco Fontana specillis a se inventis, et ad summam perfectionem perductis, editae*" (1646) was published in Naples by Gaffaro in 1646. The title makes an explicit claim that he was the inventor of the instruments used for the observations and the same claim was iterated quite obsessively in several passages of the book. Dedicated to Cardinal Camillo Pamphili, the book opens with four testimonies supporting this claim. The first one is from the Jesuit Giovanni Battista Zupus (1589-1667), who had been professor of mathematics in the Jesuit College in Naples for 27 years. In his declaration father Zupus asserted that he had first used Fontana's telescope in 1614 together with his master Jacobo Staserio, and that through his own direct observations he could confirm all the discoveries announced by Fontana:

*I, Jo. Baptista Zupus of the Society of Jesus in the kindly Neapolitan College, Professor of Mathematical Sciences, assert that many, if not all the phenomena, which Dom. Francesco Fontana is bringing to the public domain in print, not once or twice but on several occasions by me and by others of our Society by means of the very optic tubes constructed by the same Dom. Fontana…I assert that he was he who first employed two convex lenses in optical tubes, beginning in the fourteenth year of this century when he displayed for inspection a tube equipped with such lenses both to Jacobo Staserio, my Master, and to me, to the surprise and delight of us both* (translation by Peter Fay and Sally Beaumont 2001).

A second declaration by Gerolamo Sirsalis stated that Fontana invented both the telescope and the microscope. The remaining testimonies are two eulogies, one from an anonymous scholar and the other from Ippolito Vigiliis, a monk in Cassino, reader in Philosophy at the cloister of St Severino in Naples and member of the Academia degli Oziosi. The latter in particular supported the truthfulness of the numerous discoveries made by Fontana, and also stated that he had made his own telescope, though he failed to mention a date.

The *Novae Celestium observationes*, as we will call it for short, contains an etching with the author self-portrait, which is shown in Fig. 1. The oval framework holds the inscription: "*Franciscus Fontana Neapol. novi optici tubi astronomici inventor A. Dom. M.DC.VIII Aet. suae 61*" where Fontana identified himself as the telescope inventor at the age of 61. The age reported could either be Fontana's age in 1646 which implies that he was born in 1585 and invented the telescope at the age of 23 or, as suggested by Favaro

(1903), it could be read in reverse as 19, which implies that Fontana was 19 years old when invented the telescope and born in 1589. Both hypotheses are consistent with the range of the possible Fontana's year of birth.

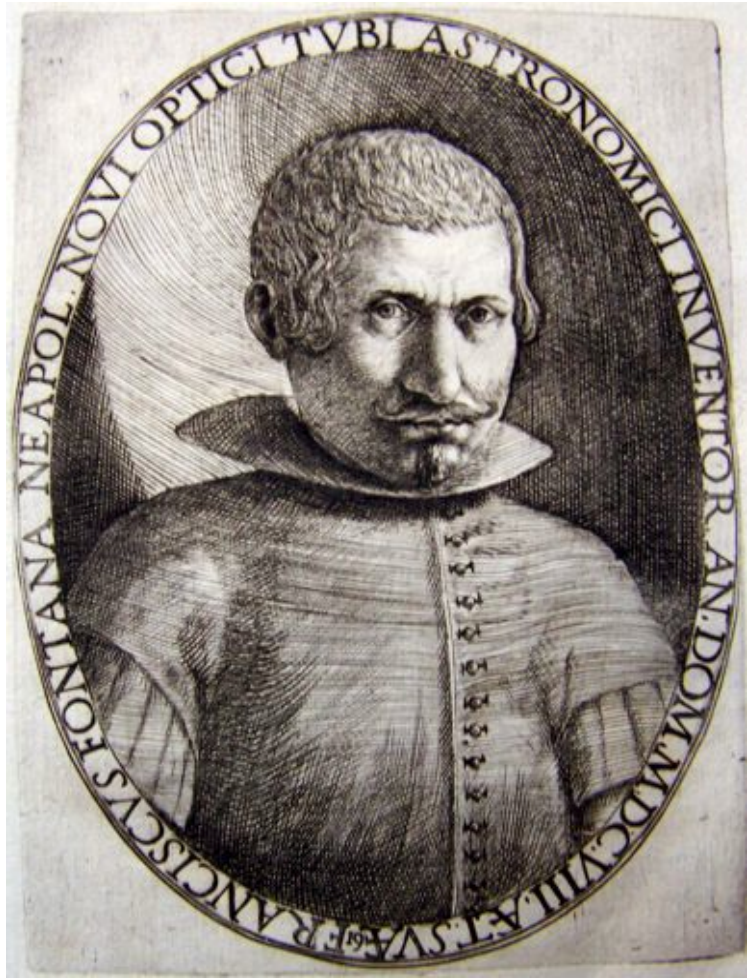

Fig. 1. Engraving of Fontana self-portrait printed in the Novas Observationes. The oval framework holds the inscription: "*Franciscus Fontana Neapol. novi optici tubi astronomici inventor A. Dom. M.DC.VIII Aet. suae 61*" where Fontana identifed himself as the telescope inventor. (Source of all reproductions is the volume of Perkins Library of Duke University).

In the "*Prefatio ad lectorem*" after recalling once more that he had invented an optical tube made of two convex lenses "*Tubi quadam optici a me anno 1608, duobus lentibus conuexis compositi inventione reperta*", Fontana explained the motivations behind his book. He complained that various authors such as Michael Florentius Langrenus or Athanasius Kircher had circulated papers based on his planetary observations without crediting him. The only exception was George Polacco who in his "Catholic Treatise against Copernicus" gave credit to his claims. An extensive list of these forged copies, including some not mentioned by Fontana himself, is given in Van de Vijver (1971a,b). Fontana explicitly said that to avoid that "*others reap the glory for themselves of all my hard work... I wished to collect quickly everything together*".

The *Novae Celestium observationes* contained observations Fontana made since 1629, along with more recent ones he mostly performed in the last two months of 1645 and the beginning of 1646. However,

Fontana considered the material presented as incomplete and warned *"I could not finish for lack of health and time"*.

The book is structured into eight treaties: the first one is dedicated to the telescope; the following three to his observations of the Moon; the fifth to the planets Mercury and Venus; the sixth to Mars and Jupiter; and the seventh to Saturn and the Pleiades. The last treaty is on the microscope. The observations are accompanied by twenty-seven full page etchings of the Moon, a larger folded plate of the full moon, and mostly full page twenty-six woodcuts of the planets, which were made by Fontana himself as he declared in the preface. This was the first atlas of the Moon, where Fontana featured images of our satellite at nearly every phase of the lunar cycle. A sort of illustrated astronomical book which was to become very popular at a later time (cfr. Winkler and Van Helden, 1992).

## 3. Tractatus primus: De Tubo Optico [4]

The first book entitled *De Tubo Optico* was entirely dedicated to the telescope. Fontana believed that the telescope had been first theorized by Giovanni Battista della Porta and then put into practice and refined by Galileo [5]. Fontana also included the verses of the Lyncean Johan Faber, doctor and herbalist of the Pope, who celebrated Galileo as the first scientist of his times [6]. This was quite noteworthy as Fontana was close to the Jesuits of Naples notably hostile to Galileo and whose permission was needed for him to publish. Considering Porta as the inventor of the telescope in 1589, makes clear that Fontana's claim to have invented the telescope in 1608 referred exclusively to the telescope made by combining two convex lenses.

Further on in his book Fontana provided a brief excursus on the instrument's history from antiquity to his age. Fontana rejected the possibility that the ancients already knew the telescope on the grounds that they had never revealed any detail of the Moon and the stars. All important discoveries about planets and stars by means of a telescope had been made by Galileo. A detailed list followed: i) the Milky way was made of stars, ii) the *hazy* stars were composed of multiple stars, iii) the number of fixed stars was 10 or 20 times that given by Ptolemy, iv) Jupiter had four satellites, v) the Moon was not a perfect sphere, vi) Saturn was consisted of three stars and vii) Venus had phases. After Galileo the only significant discovery had been made by Langrenus in 1645, with his map of the Moon showing dark spots. Langrenus was the first to put forward a system of lunar nomencalture with few names that survive till today. However, Fontana added that this map could have been "*derived possibly from my maps*… first done in 1629" "*since Langrenus never reveals the designer of his telescope*".

Fontana wrote that with his own telescopes he had confirmed all those discoveries, i.e. in an apparently "*empty*" sky with the telescope there are in fact "*now 3 now 4*" stars, the Pleiades were made at least by 28 stars, the Nebulae were composed of stars and the Milky Way by an infinite number of them.

The difficulty of working lenses in order to give them a perfect spherical shape was then described, including the role played by bubbles and air-holes in the glass. He stressed the importance of possessing a testing tool to check for the lens-shape, and he proposed to look at the projected image of a candle as a testing procedure for the lens quality (Fontana called this his first invention). In a chapter entitled "*Concerning the Astronomical telescope invented by the author*" the construction of the author's second invention is described. Fontana clarified that when he had conceived the idea of his telescope, he did not know Kepler's Dioptrice (1611):

"*Although that model seems to be proposed by Johann Kepler in his Dioptrics, Question 86, p42 printed in 1611. However, I had in truth no knowledge of this book earlier than the present moment when I am publishing this treatise, and I have received it in return from the aforementioned Johan Baptiste Zupus. It is surprising that it is not recorded that Kepler was the inventor of this device in Germany and myself in Naples also his method is quite different from the method suggested here, read it* (p 20 of Novae Observationes, translation by Fay and Beaumont 2001).

In the last sentence Fontana seems to doubt the real intentions of Kepler and invites the reader to go directly to the source: *legite ipsum*. Fontana described also how to correct inverted images by the use of a third lens with the same radius of curvature (his third invention), apparently ignoring the presence of a similar concept in the Dioptrice by Kepler and in *Oculus Enoch et Elliae Sive Radius Radius Sidereomysticus* (1645) by Anton Maria Schyrleus de Rheita. An astronomical and terrestrial telescope thought to have been made by Fontana around 1650 is at the Museum of Optics by Luxottica in Agordo. It is a terrestrial telescope with an eyepiece composed by 3 lenses and could be an early implementation of Fontana's third invention.

The last chapter described how to construct very long telescopes, i.e. with a length up to 50 palm, 13.18 m, as the Neapolitan palm corresponded to 0.2637 m. With such a length the radius of curvature of the lenses is so large that the lenses surface became almost flat and therefore extremely difficult to work. Fontana described his solution to the problem by introducing for the first time the concept of optical meniscus *"This inconvenience will be avoided, if the glass is figured on one side in a convex shape and on the other side in a concave one"*. Fontana considered this his fourth invention. But he does not mention the problem with chromatic aberration which severely affect these kind of telescopes.

### 3.1 On Fontana's telescopes

Some information on his telescopes can be obtained from the correspondence between the natural philosophers and scholars of his times. A first mention was contained in the letter of Fabio Colonna to Federico Cesi of 30 November 1629: *il sig. F. Fontana ... ha fatto un cannone di otto palmi [2.1m], con il quale se ben allo rovescio fa vedere la luna et stelle ..."* In 1637 Fontana, while trying to sell his telescopes to the Grand Duke of Tuscany Federico II, contacted Benedetto Castelli, who wrote to Galileo celebrating the virtues of Fontana's telescopes [7].

In the following year, Fontana improved his telescope by making a 14-Neapolitan- palm-long (i.e. 3.7 m.) telescope. The construction of this very long telescope was documented by a letter of G. G. Cozzolani to C. A. Manzini of 11 Sept 1638, as well as by two letters that Castelli wrote to Galileo in July 1638. On 3 July, Castelli wrote: *I am holding a glass of Naples that is for a telescope long fourteen Neapolitan palms, [] magnifies the object ninety times* [8] and, a few days later, on 17 July the magnification became *160 times .. a monstrosity*. This telescope was then bought by the Extraordinary Imperial Ambassador in Rome, the duke of Cremau, Prince Ecchembergh (Del Santo 2009)

Fontana's grinding and polishing technique still remains unknown as it was only partially disclosed in his book. On 3 January 1638 Fontana wrote to the Grand Duke of Tuscany with the offer to reveal his secret way to work lenses for a reward of 2000 piastres but the Grand Duke declined the offer. This attempt is also recorded also in a letter of Castelli to Santini in the same year (cfr Arrighi, 1964). In a letter of 10 July 1638 Castelli wrote to Galileo that he thought he had figured out Fontana's secret way of grinding lenses. Apparently Fontana was working only the central part of the lens, as we deduce from Galileo's answer of 20 July 1638 [9].

On 23 October 1639, Fontana directly addressed the Grand Duke of Tuscany proposing a 22-palm-long, i.e. 5.8m, telescope (cfr Paolo del Santo 2009). We do not know the Grand Duke's answer, though del Santo suggests that this telescope had actually reached Florence.

### 3.2 The genesis of the astronomical telescope

Four centuries later the details of the genesis of the Dutch telescope are still unknown, but even more mysterious is the birth of the astronomical telescope, i.e. the one made up with two convex lenses (Van Helden 1976, Van Helden 1977a,b, Van Helden et al 2011), . After the publication of the *Sidereus Nuncius* by Galileo, in the summer of 1610, Johannes Kepler wrote *Dioptrice*, whose publication followed one year later. Kepler's book was devoted to the explanation of the Dutch telescope functioning but also considered all other possible combinations of lenses, including two and three convex lenses. However, these considerations were not inserted in the Dioptrice's section on the telescope and, when discussing the image formation, Kepler did not mention the magnification which is the main characteristic of a telescope. As argued by Malet (2010) *"the idea of turning his theoretical combination of two convex lenses into a working telescope may have never crossed Kepler's mind"*. A similar doubt was expressed by Fontana, when invited to read carefully the Kepler's book. As a matter of fact Kepler did not make a telescope, and we had to wait as long as 1645 when De Reitha manufactured the first *"Keplerian"* telescope, apparently on the basis of Kepler's instructions.

In 1618 a claim for primogeniture in the construction of a *"long tube"* was made by Johannes Sachariassen in favour of his father Sacharias Janssen in an answer to the Middelburg City Council investigation that had been set up in 1655to clarify the origin of the telescope [10]. However, several inconsistencies were noted in his declaration (cfr Van Helden 1976, Zuidervaart 2011) and probably the definition of *"long tubes"* did not refer to a Keplerian telescope but to a Dutch one with a longer focal length (Van Helden 1976,). TH

The first printed mention of a telescope formed by two convex lenses appeared in *"Rosa Ursina sive Sol"* (1631) by Christoph Scheiner.When describing how to use a Dutch telescope to project the solar image, he mentioned that a different arrangement for the projection which made use of two convex lenses was also possible [11]. At page 130 he wrote: " *thirteen years ago, I made erect the images intercepted for the most Serene Maximilian, Archduke of Austria"*. Thirteen years before the publication date was the year 1617; but since *Rosa Ursina* took a four-year period to be printed, it could have been within 1613-1617 (cfr Van Helden 1976). However, a document of 1616, kept in the Tyrolean State Museum Ferdinandeum, states that " *the Archduke [Maximilian] acquired an optical instrument of admirable utility but that was giving inverted images; since he wished to see the pictures erect, and this could not be obtained he turned to the Jesuits, who gave him the Professor of Mathematics in Ingolstadt [Christoph Scheiner] as an expert* [12] (Daxecker 2004). This document, which is the first document to refer to an astronomical telescope, confirms Scheiner's reconstruction and fixes the date at 1616. However, neither this document, nor the Rosa Ursina did mention Scheiner as its inventor. He was simply reported to have added a lens to a pre-existing telescope and rectified the image for the benefit of Maximilian III. Moreover, neither in *Disquisitiones Mathematicae* (1614), nor in the manuscript *Tractatus de Tubo Optico* (1616), and not even in *Oculus hoc est fundamentum opticum* (1619), has Scheiner ever referred to himself as the inventor of a Keplerian telescope. An omission that would be very bizarre if he really were the inventor of a new kind of telescope. So very little was known of Kepler's telescope that, when Antonio Maria Schyrle de Rheita mentioned it in his *Oculos Enoch et Eliae* (1645) – disregarding altogether the contribution of Fontana – he was generally credited with this invention (cfr King 1955). An opposite attitude is that of Francesco Fontana who in 1646 claimed throughout the book the primogeniture of the construction of a telescope made with two convex lenses. There are no apparent reasons to question Father Zupus's declaration to have used Fontana's telescope in 1614, since he was still alive when the book was published. Allowing some time to improve the quality of the lenses even the year of 1608 does not seem so implausible as the birthdate of Fontana's telescope, though it is based only on his own words. The improvement in the optical quality of the lenses has been probably the decisive factor if we consider that already in 1538 the Italian scholar Girolamo Fracastoro wrote: *"If someone looks through two eye-glasses of which one is placed above the other, he shall see everything larger and closely"*.

## 4. De Lunae Observationibus

Fontana dedicated 3 books of his *Novae Celestium Observationes* to the Moon. The first is a summary of all his discoveries on the Moon; the second presents 13 observations of the waxing moon, and the third reported 11 observations of the waning Moon made in January 1646 together with four previous lunar observations made in 1629, in 1630 (two) and in 1640. Fontana considered the results of these earliest observations as less accurate since *"they took place at a time when the optic tube had not reached its present standard of perfection"* and presented them at the end of the fourth book. They had probably been obtained with his telescope of 8 palms while, for the last observations, he probably used the 12-palm one. However, his earliest observations are more interesting to us since they are the first observations ever performed with an astronomical telescope. Inverting Fontana's original order, we have considered these ones in the first place.

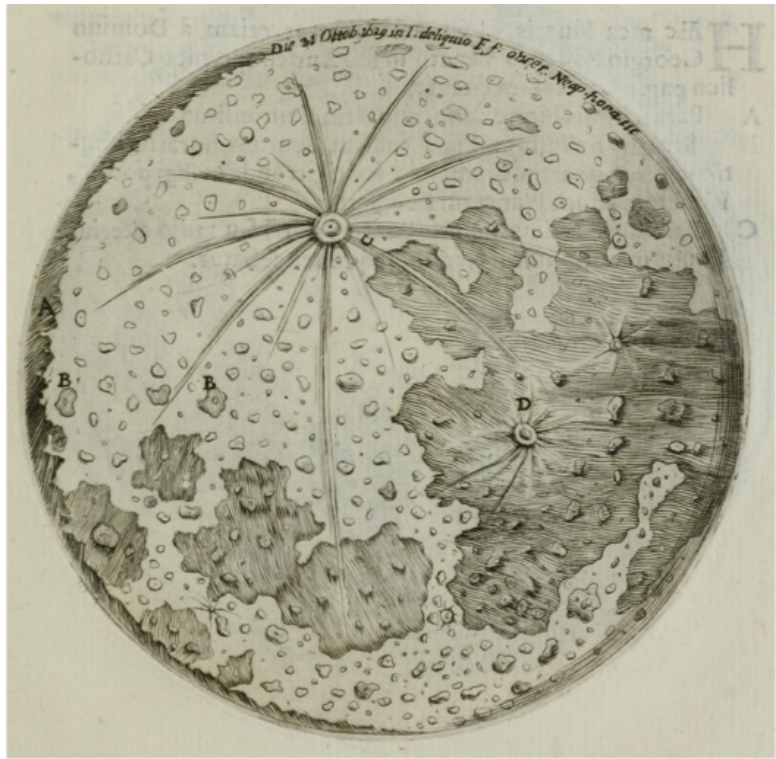

Fig. 2. Moon observation of 31 Oct 1629 taken 3 hours after the sunset (*hora ad occasu Solis tertia*). The etching has a size of 10.3 cm. The Moon is upside-down with the South on top and East to the right as seen with an *astronomical* telescope (source Perkins Library of the Duke University). Some features in the picture are marked with letters: A) highlighted that the Moon was not perfectly spherical but irregular at the border as an axe blade. Fontana wrote "Terminus corporis illuminati adhuc cernebatur tunc no perfectus circulus, sed inaequalis, securi similis" and he has been the first one to see the irregular shape of the Moon. We recall that in the Sidereus Galileo, after postulating the presence of mountains and had measured their heights, was surprised not to see an irregular border and had postulated the presence of an atmosphere. B) indicated a new though relatively small spot. Letter C) indicated what we know today as the crater Tycho. This crater has been observed (in this position) for the very first time by Fontana, who also saw several rays formed by the materials splashed out by the impact a well as the central peak, which is a characteristic of big craters. Fontana named it *Fons Major*, i.e. biggest fountain, echoing his name Fontana which in Italian means fountain. Letter D) marks the crater today known as Copernicus, which was also seen for the first time by Fontana.

## 4.1 Moon observations of 1629 and 1630

Fig 2 shows the Moon observation of 31 Oct 1629 made three hours after the sunset, probably from the roof of Fontana's house in Naples. The quality of this map can be judged by comparing it to those available in the same years, for instance C. Malapert's (1619), G. Biancani's (1620) and C. Borri's (1627), as well as in comparison with a modern high resolution image of the Moon as shown in Fig 3. (cfr. Whitaker 1999). Definitely, Fontana was the first one to draw the true shape of both the Moon's markings and major craters.

Fontana's etchings of the Moon had circulated all around Europe long before they were published in the *Novae Celestium Observationes*. For instance, one of his lunar maps was sent to scholars in Genoa by Castelli, as documented in the letter by Renieri to Galileo on 5 March 1638: *A picture of the Moon is arrived in Genoa, sent here by Benedetto Castelli, with news of a new telescope invented by a certain Fontana from Naples showing things more exquisitely than any other* [13]. From a letter Gloriosi wrote to Santini on 13 March 1638, we learn that copies of the Moon's maps had actually reached Cardinal Barberini and the Grand Duke of Tuscany, as well as other persons [14] (cfr Arrighi 1964).

Fontana's lunar maps were reproduced by several authors. Hirzgarter used them in his *Detectio Dioptrica Corporum Planetarum velorum* (1643), Argoli in his *Pandosium Sphaericum*, (1644), father Athanasius Kircher in De arte magna de lucis et umbrae (1646), and Giorgio Polacco in his *Anticopernicus Catholicus*, (1646) (van de Vijver 1971). Fontana suggested that his maps might have been the source of the map of Langrenus, the Royal Cartographer of king Philip IV of Spain who, in 1645, had provided a first nomenclature of the Moon features, some of which are still in use today. As Fontana explained in his opening address to the reader, the wish to protect the paternity of his work was one of the motivations to write his book.

The observation of 20 Jun 1630 shown in Fig. 4 is of special interest since it recorded a rare occultation of Saturn by the Moon. Fontana wrote that the occultation took place on 20 June 1630, started about 3 hours after sunset and lasted less than two hours. Actually, the occultation took place on 19 June, one and a half hour after sunset and lasted less than one hour. These differences have often been considered as the evidence of Fontana's overall inaccuracy. However, these inaccuracies can be explained if we consider the way Fontana recorded his observations. As suggested by Beaumont and Fay (2001) Fontana takes the sunset as the start of the day instead of midnight. Thus, the 3rd hour after sunset of 20 June corresponded to the evening of 19 June. In the Skygazer simulation of the event the occultation from Naples started at 22:10:19 (UT) and ended at 22:58:59, for a total duration of about 49 minutes., thus much shorter than the two hours reported by Fontana Nonetheless, I suggest that Fontana could possibly have adopted a Roman timekeeping. In this system there are 12 hours between sunset and sunrise and the length of the hour over the year and between night and day is variable. The occultation took place almost at Summer solstice, when the night hours are shortest. Adopting a modern astronomical definition of sunset, the end and the start of twilight at the time in Naples were respectively at 20:44 UT and at 1:23 UT, which yields an hour length of 4:39m/12h = 23,25 minutes. It is possible that Fontana used a less strict definition for sunset, such as the civil or the nautical sunset, with the Sun at 6 or 12 degrees, respectively, thus getting an average hour slightly longer than 25 minutes, thus accounting for the length reported by Fontana Moreover, this hypothesis provides consistency also to the start of the occultation which he said to have occurred 3 hours after sunset. With sunset at 20:44 UT and a 28-minute-long hour, the occultation started around 22h:08m UT, in agreement with the Skygazer simulation of 22h:10m (UT). Fig. 4 reproduces start and end of the occultation of our simulation, as well as Fontana's drawing, showing that the positions had been accurately drawn in the etching.

The third old etchings of the Moon referred to an observation of 24 June 1630, just few days later than the previous one. Here Fontana noted that the Chief Fountain (Tycho) was nearer to the centre of the Moon, which implied that the Moon was rocking back and forth. In the fourth observation of 9 June 1640 Fontana noted that the Chief Fountain (Tycho crater) was closer to the centre of the Moon than he had ever seen. Moreover, the middle point of greater marking (Oceanus Procellarum) was at the border, definitively proving the existence of the third motion, i.e. the E-W motion. It would be really remarkable if Fontana really had made these considerations already in the Summer of 1630.

Before presenting his observations, Fontana summarized his lunar discoveries in Book II. The first chapter opens with a theory on the source of the Moon's light that he believed to come from the Sun, though, always according to him, a feeble light originated also from the Moon itself which could be seen in the non illuminated part. The origin of the secondary light was a quite controversial issue with Galileo defending the interpretation of his terrestrial origin and the Jesuits the opposite view (cfr Molaro 2013). In the second chapter, the Moon's shape was discussed and reported as irregular: *"A large number of observations seem to indicate that the Moon is not a perfectly spherical body, but on its surface various irregularities are to be found"* (cfr note A in Fig. 2). Chapter III described the lunar markings, whose irregularities are considered to have the same nature of the other ones. Chapter IV was dedicated to the lunar movements. As already noted, the observation of the *Great Fountain* revealed a North-South direction movement which was to be added to the already known swift in East-West direction. Galileo first described the Moon's libration in depth in a letter of 7 November 1637 and returned on the subject in a letter to

Alfonso Antonini of Udine on 20 February 1638. This letter, which Galileo had asked to keep reserved, was published only in 1656, in the Bologna edition of Galileo's works, i.e. after the publication of Fontana's book. It seems therefore unlikely that Fontana had read Galileo's private letter. Fontana connected the Moon's motions with the rotation which he assumed to last 27 days, as the solar rotation estimated by C. Scheiner in *The revolution of the Sun* (Book IV, Part II, Chapter 10) of the *Rosa Ursinae*. It is interesting to note that Fontana argued that the Moon rotation and the N-S motion implied that the Moon could not be a fixed body onto a celestial sphere which, according to Aristotle, was moving East-West. The same argument will be used later on for the other planets which he found to rotate on their own axis.

Book III and Book IV contain 13 etchings of the waxing Moon and 11 of the waning Moon, a number meant to show how the lunar features change with phase. Fontana also remarked that he had been capable to reproduce only a thousandth part of the details that he had been able to appreciate with its telescope. Such a detailed Moon Atlas has no precedents and is the first astronomically illustrated book (cfr Winkler & Van Helden, 1992). Of particular interest is the observation N. 10 where together with a lunar observation of November 1645 Fontana summarized his main planetary discoveries in the four corners of the etching. The label in the round framework recalls that the observation had been performed with a "*Thelescopio ad ipso invento 1608*", where the word *Thelescopio* is used here for the first and last time.

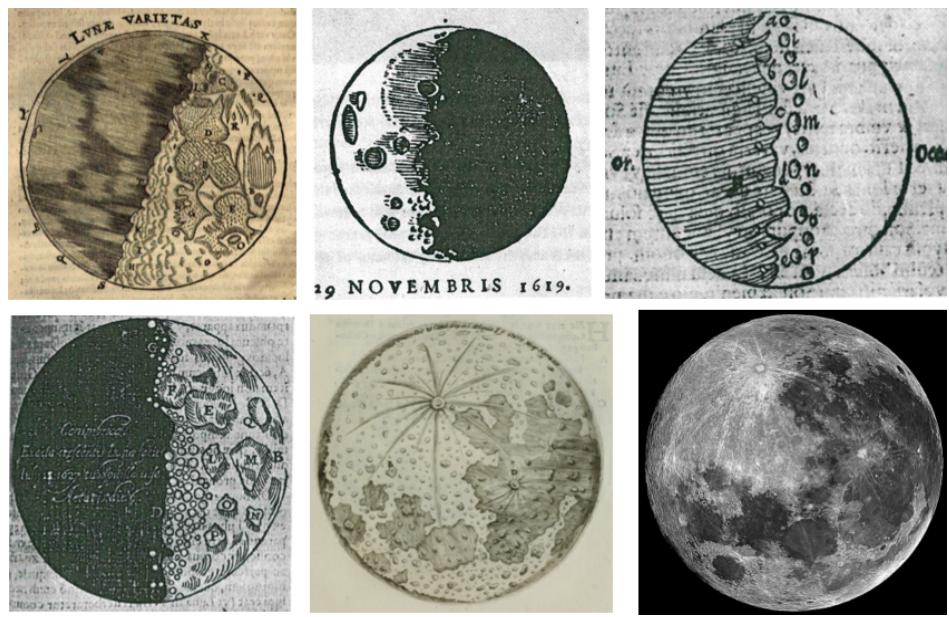

Fig. 3. Moon's drawings before Fontana's observations. From top left to right: Christophe Scheiner (1614), Charles Malapert (1619,) Giuseppe Biancani (1620). Bottom from left to right: Christopher Borri (1627), Francesco Fontana (1629) upside down, modern view also shown upside down.

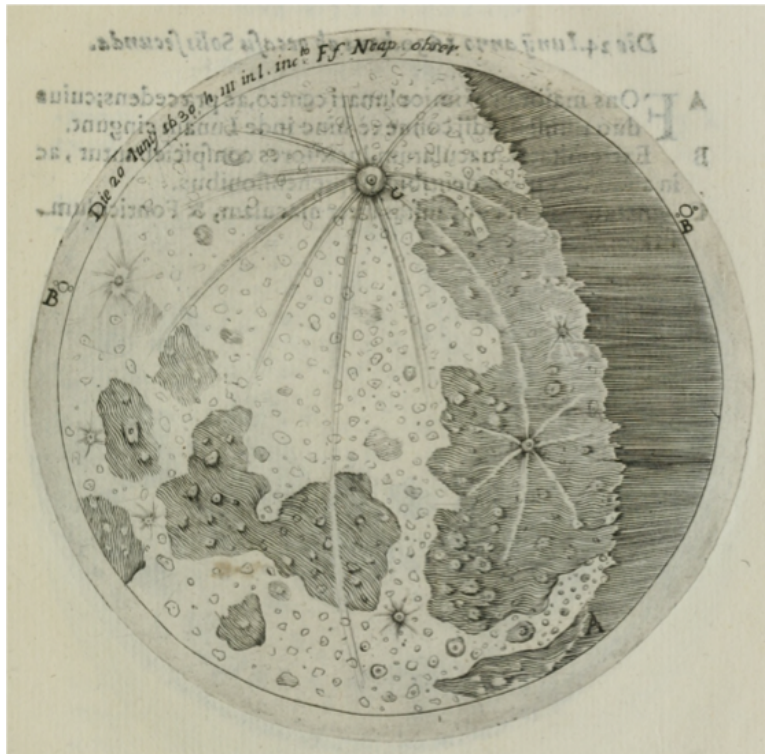

Fig. 4. Observation of 20 June 1630. Diameter of 10 cm. The inscription around the Moon reports: Die 20 Juny 1630 h III in l. inc.to. F.f. Neap obser. Beside letter B there is the Galilean symbol of Saturn made with three stars marking the position of Saturn both at the start and the end of the occultation. Letter A highlights the presence of a special darker area than the dark surrounding and letter C a stream of rays that from the Chief Fountain (Tycho) join up with a ray originated from the other great fountain (Copernicus) in the great dark spot (Oceanus Procellarum).

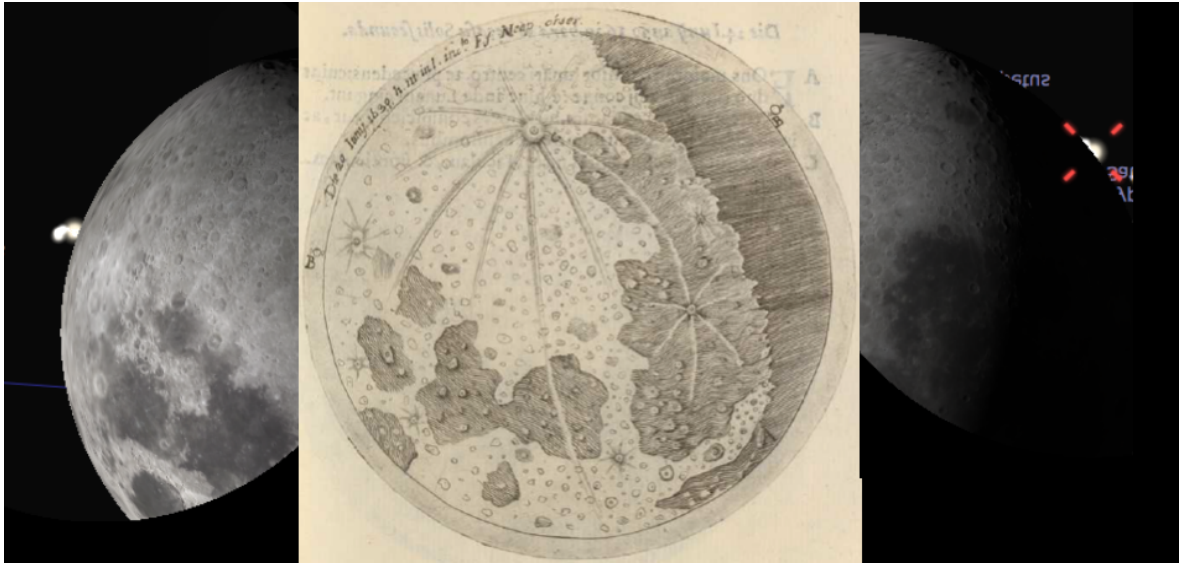

Fig. 5. Same as the previous figure with the end of the occultation on the left side and start on the right side. The eclipse is reconstructed with Skygazer 4.5. This shows that the positions marked by Fontana are drawn precisely.

## 5. Tractatus Quintus. De Mercurij, & Veneris Observationibus

### 5.1 Mercury observations

Two observations of Mercury are presented in book 5 in which the planet is described as *"curved like a bow with the concave edge pointing towards the sky and the convex edge turned towards the horizon"*. Thus Mercury revealed its phases conclusively showing that it is orbiting around the Sun. Fontana revealed that these observations had not been made by him but by father John Zupus with one of his telescopes. In Fig 6 we show the woodcuts of the two Mercury observations together with the Skygazer simulations of the planet seen from Naples in the dates provided. The results of the simulations are giving a percentage of illumination of around 40% on 23 May 1639 and of around 36% on January 1646, quite consistent with the drawings.

In his *Almagestum Novum* Riccioli ascribed the former observation to Zupus and the latter to Fontana. He himself observed Mercury's phases in the years 1643-1644 and 1647. Riccioli considered the detection of Mercury's phases a very difficult observation because of the small dimensions of the planet *("à Sole vaporesq. horizontis, rarò dato opportunitas eum falcarum videndi"*, Almagestum Novum lib. VII, sectio I, cap. II, p. 484). These observations show both the quality of the telescope and Fontana's rectitude in attributing the discovery to father Zupus.

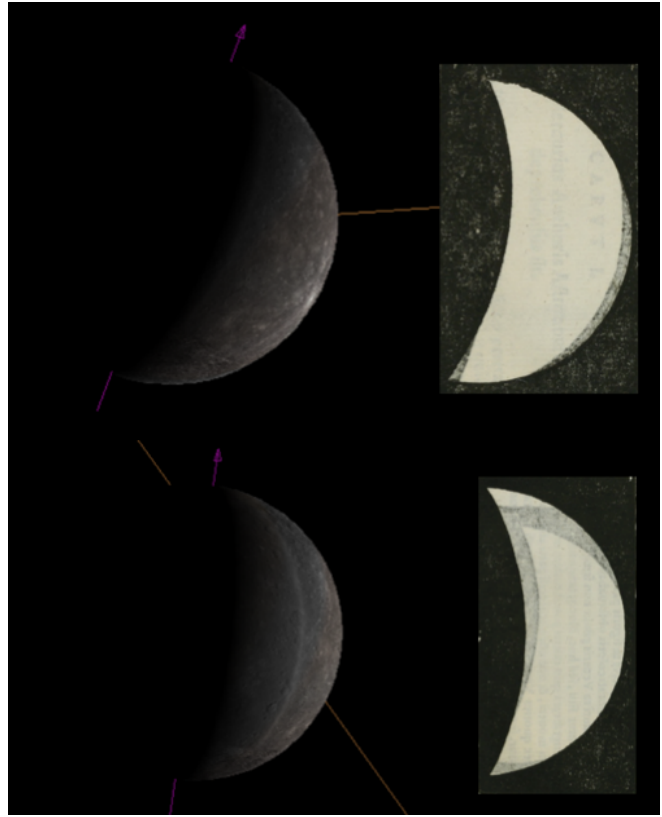

Fig. 6. Top: Mercury as seen from Naples on 23 May 1639. The concave edge pointing towards the sky and the convex edge turned towards the horizon. Bottom: Mercury on January 1646. The cusps of Mercury's concave side pointed to the sky at a different angle than in the first observation.

## 5.2 Venus observations

Six observations of Venus are shown in six drawings. The first was made on 22 January 1643 and the last on March 1646, probably the last observation recorded. They show Venus's phases at their best and are reproduced in Fig 7. The simulations with Stargazer are also shown beside the pictures providing an illuminated fraction of 17% and 35% in good agreement with Fontana's drawings. Fontana also noted that the concave side showed an irregular edge with the light appearing a little dimmer near the edge, a phenomenon known as *terminator shading*. In particular, using these two observations Fontana thought that Venus had an oval shape and that therefore the change in its appearance implied it was rotating around its axis.

The remaining four observations, obtained between 11 November 1645 and 22 January 1646, besides confirming the phases also reported the presence of two moons. *This is a new discovery not yet published in my opinion. But it is true that they do not always appear, but only when Venus is shimmering.[] These little dots were...not always seen in the same situation on Venus, but they moved back and forth like fish in the sea.* This claim originated a controversy that lasted for more than one hundred years. A detailed account of this research has been provided by Kragh (2008). Riccioli said he had never observed the moons and Huygens in 1659, after 3 years of observations, concluded that there were no moons. On the other hand, Cassini saw a moon in 1672 and in 1686, but never again. Short saw a luminous object close to Venus on 3 November 1740 and A. Mayer on 20 May 1759. The issue was resolved only when the Venus transit of 1761 did not show any moon. It is a note of curiosity that in the same year Venus moons were seen 19 times. An explanation in terms of optical reflections in the telescope was published in the *De satellite Veneris* (1765) by the Jesuit M. Hell and in 1881 W. F. Denning provided a similar explanation.
It is very likely that Fontana's telescope was affected by some light reflection particularly evident when observing a bright object such as Venus, responsible also for the reported presence of rays.

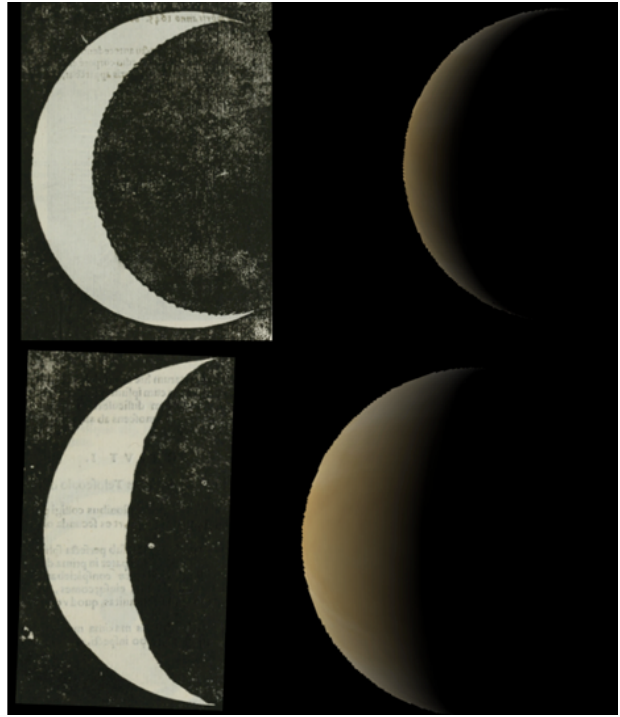

Fig. 7. Top: Venus on 22 January 1643 hora ad occasu Solis secunda. The illumination was of only 17% matching very well Fontana's observations. Bottom: March 1646. The illumination is of 35% as in the drawing. Fontana notes also that the brightness was unequally distributed with the light appearing dimmer near the concave edge, an effect known today as terminator shading.

In fact, Fontana in the comments on the third observation of 15 Nov 1645 noted that "*Two starlike points of that same subdued reddish colour were seen, one at each of Venus' cusps, almost adjoining them. Although this appearance of Venus, if it is a sphere and receives its light from the Sun, might be an optical illusion, yet this is how it really-looks*". And in the comment of the fifth observation he said a little globe or spot was facing the concave edge of the *real* Venus. Where the word real is literally 'more true', and Beaumont and Fay (2001) commented that Fontana suspected that the little globe could be an optical illusion. In retrospect, it seems that this wrong prediction had a particular weight in the negative judgement reserved to Fontana by astronomers of all times.

## 6. Tractatus sextus: de Martis & Iovis observationibus

### 6.1 Mars observations

Fontana observed a gibbous Mars with a *black cone* like a hollow in the middle planet. This was probably the Syrtis Major, a mark established few years later by Christian Huygens and Robert Hooke. Fig 9 shows Mars observations of 1636 without date and the one of 24 August 1638. While in the former there is no evidence of phase, the latter shows a gibbous Mars. This feature of Mars was also seen by Benedetto Castelli with a Fontana's telescope as recorded in his letter to Galileo of 17 July 1638. At that date the illuminated fraction of Mars was of 83.3% as the drawing suggests. The dark spot at the center changed quickly and since the hour was not provided we have not attempted to reproduce its position. Note that Fontana reported also the presence of a ring which does not exist. From the quick motion of the dark spot Fontana deduced that Mars revolves around its own axis. The book is not very clear on this point where in both drawings the dark spot is approximately in the same position. However, a letter from Cozzopani of 11 September 1638 to Carlo Antonio Manzini revealed some of the discussions induced by Fontana's observations much in advance of their publication: *"in the Mars's center there is a prominence as a black velvet ending in cone shape and around there are two circles or two bands ... and everything is mobile, since you do not look in the same place .."* [15] On 17 July 1638 Castelli wrote to Galileo to have seen a gibbous Mars with a Fontana's telescope. Three days later Galileo answered to say how beautiful was this observation. In a letter of 15th January 1639 to an unknown Galileo wrote: *"As to the planet Mars it was observed that being at the square with the Sun, it is not seen perfectly round, but rather flared, similar to the Moon of 12 or 13 days, which from the side opposite to that touched by the solar rays it remains unilluminated, and consequently not seen, what I have said should have happened when Mars was seen superior to the Sun"* [16]

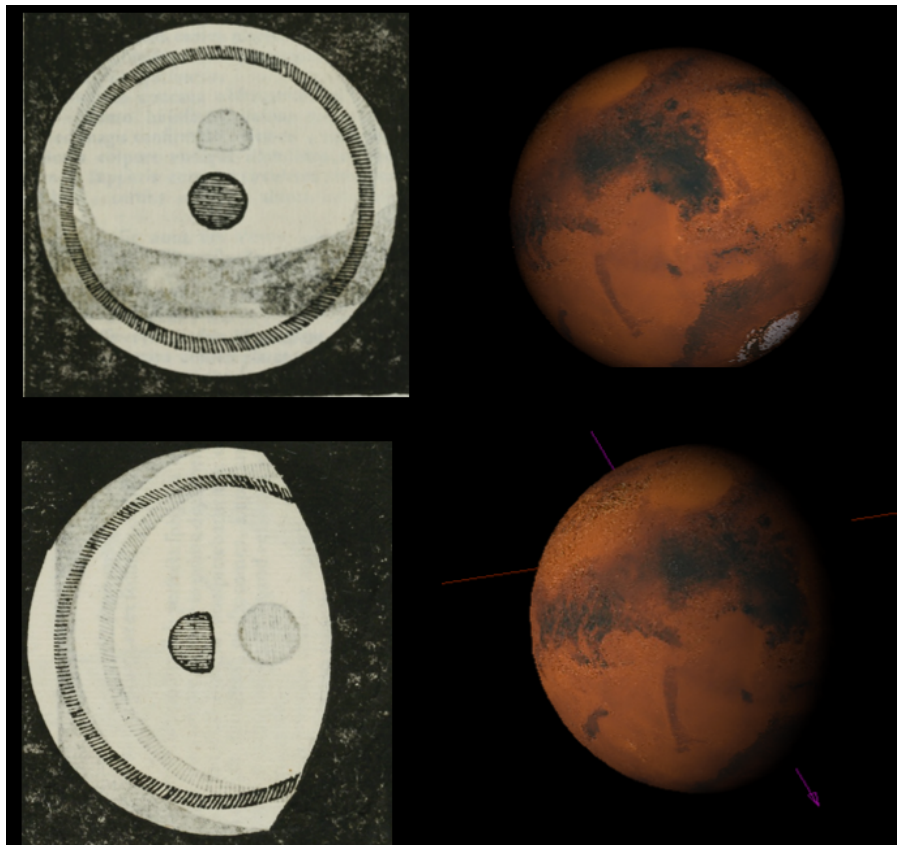

Fig. 9. Mars observations of 1636, without date, and the observation of 24 August 1638. The former is perfectly round, the latter shows a gibbous Mars. Next to the right the simulations with Stargazer are shown. In 1638 the illuminated fraction of Mars is of 83.3% as the drawing suggests. The dark spot at the center changes quickly and it cannot be reproduced without the hour. Note that Fontana reported also the presence of a ring which did not exist.

## 6.2 Jupiter observations

Fontana presented eight observations of Jupiter ranging from 1630 to 1646. The planet was found perfectly spherical but on the globe he noticed, already in 1630, some bands which persisted in separate observations. The observations of the bands were confirmed by Zupus independently with a different telescope. Fontana himself observed the bands with different telescopes to be sure of their existence.

Fathers Niccolo Zucchi (1586-1670) and Daniello Bartoli (1608-1685) are sometimes credited for having seen Jupiter's bands also in 1630, but there is no proved documentation and we think that the sources are Fontana and Zupus' observations. Indeed, only they could avail of an instrument capable of observing the Jupiter bands. According to a letter by Torricelli of 10 Feb 1646, Castelli saw Jupiter bands in 1632 from Rome (cfr Paolo del Santo 2009).

The bands were "*not more than three not fewer than two*" and sometime they were seen as convex curves, sometime concave, and also as straight lines. The bands are circular clefts with some hollow spots on them. From the change of shape of the bands Fontana deduced that the planet was revolving around its own axis, and the rotation implied that the planet had an independent existence and was not attached to the revolving heavens. The new features on Jupiter itself implied a flaw in the perfection of the Aristotelian skies. When in 1639 Fontana approached the Grand Duke proposing a telescope of 22 palm, i.e. 5.8m, as a demonstration of the superiority of his telescope he attached a watercolour with the discovery of the bands on Jupiter, which are shown in Fig 10.

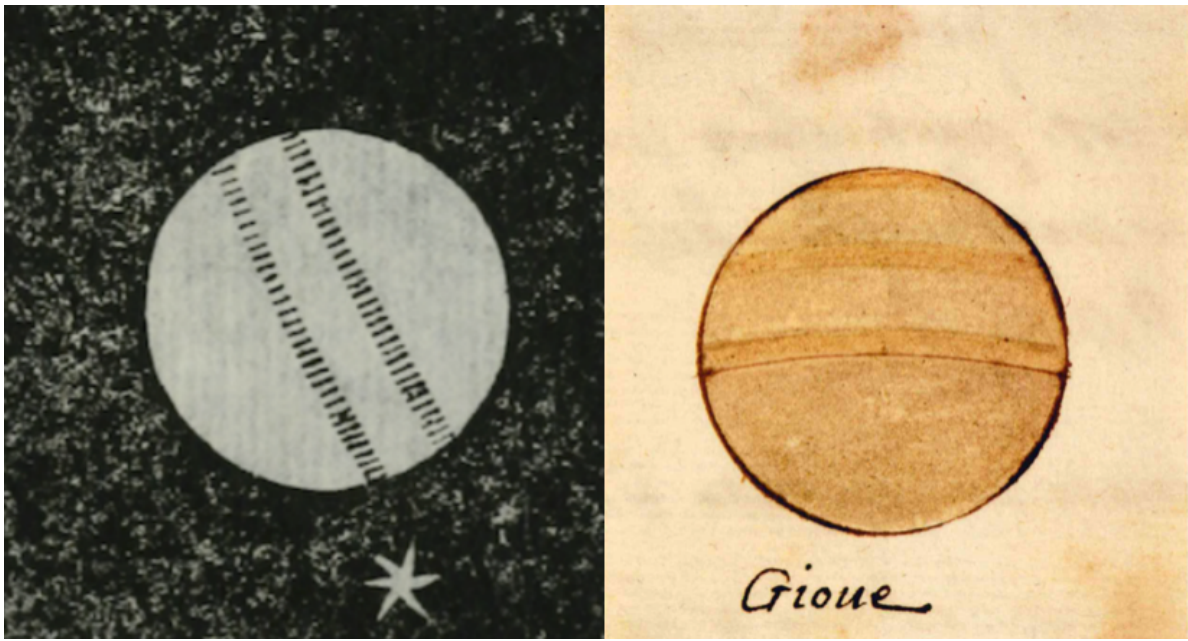

Fig. 10. Left: detail of Fontana's woodcut with Jupiter observations of 1630. The star is one of Jupiter's moons and together with the bands marks the plane normal to the axis of rotation of the planet. On the right the watercolour of Jupiter showing three bands attached to the Fontana's letter of 1639 to the Grand Duke Ferdinand II dé Medici (Courtesy Archivio di Stato di Firenze, fondo MM, busta 514,fas.1,c 64v)

In the field of Jupiter Fontana noted the persistence of 5 stars which he suggested could be moons. For an observer the argument provided by Fontana seems very naive, since it seems he did not realize he was looking at different portions of the sky:

*It can be shown that they are not fixed stars for the reason that fixed stars always keep the same positions relative to each other, as all Astronomers agree. But these are seen to behave differently. Also some fixed stars would be visible in the vicinity of Mars too, which is nearer to Jupiter in the order of the planets, and many more around Saturn, the nearest planet to the realm of fixed stars, but the opposite is the case.*

# 7. Tractatus septimus: de Saturni, & Vergiliarum observationibus

## 7.1 Saturn observations

Fontana presented a set of seven observations of Saturn which he said appeared in his telescope as the full Moon at the naked eye. The date is not always reported but they appear grouped in three groups. The first observation was made on 20 June 1630 when the planet had been eclipsed by the Moon, as we have already discussed. The drawing depicts Galileo's rendering of the planet made by three stars perfectly spherical with the middle one about two times larger than the outer ones. The second did not report the year but it looks similar to the third made in 1634. Fontana noted that the shape of the planet changed considerably. The central body is oval and the two stars seemed "*embracing the planet itself on either side*". Also in the next observations of 1636 these were seen in the form of the "*crescent moon and touching its globe*" which this time was perfectly spherical. It is quite possible that the different appearance of the planet was linked to improvements to his telescope passing from that 8 palm long to the 14 palm (i.e. from 2.1 to 3.8m). The fifth observation does not bear a date but as noted by Beaumont and Fay (2001) it should be close to the last two taken in the last observing session of 3 and 12 Dec 1645, respectively. The two satellite stars appeared more distant from the central body and "*they have on either side something in the nature of handles forming a triangular shape which seems attached to the middle of a perfect spherical body*". The observation on 3 December is very similar, but with the triangular shape of the handles becoming more oval and curved, and in those of 12 December the satellite stars are becoming smaller and more distant. The last observation of 12 December is shown in Fig 11 together with our simulation for the same day. From the simulation it is possible to appreciate how the overall proportions and the tilt of the disc were accurately drawn by Fontana. The description of the planet in the last three observations shows how close he was to reveal the real nature of the planet. Fontana's thoughts regarding the changing of the shape of Saturn can be deduced from a letter of Gloriosi of 21 September 1638, in which the observations of Fontana are commented saying that the cause is likely the change of position of Saturn with respect to the Sun. From the same we also know a that the regions within the "handles" were seen by Fontana as empty sky. The Saturn problem was solved by Cristiaan Huygens (1659) who admitted to have been inspired by Fontana's observations: *Etenim aliae deinceps mirabiles ac prodigiosae formae apparuerunt, quas primum a Josepho Blancano et Francisco Fontana descriptas novimus* (Huygens Systema Saturnium,p 535), and also "*Porro ab hisce figuris non multum recedit ea, quae a Francisco Fontana vulgata fuit, undecima tabella nostrae* (p 558).

Also for Saturn, as for the other planets and the Moon, Fontana concluded that the planet was moving freely in the sky, and therefore it was not attached to an Aristotelian celestial sphere. Around Saturn Fontana seems to have seen further moons away from the planet in several observations. As suggested by Fay and Beaumont (2001) it is possible that Fontana saw Titan and Iapetus that were relatively bright. Titan, discovered by Huygens in 1655 on 12 Dec 1645 was of 8.23 mag and separated by about 3 arcminutes from Saturn and it should have been within the reach of Fontana's telescope.

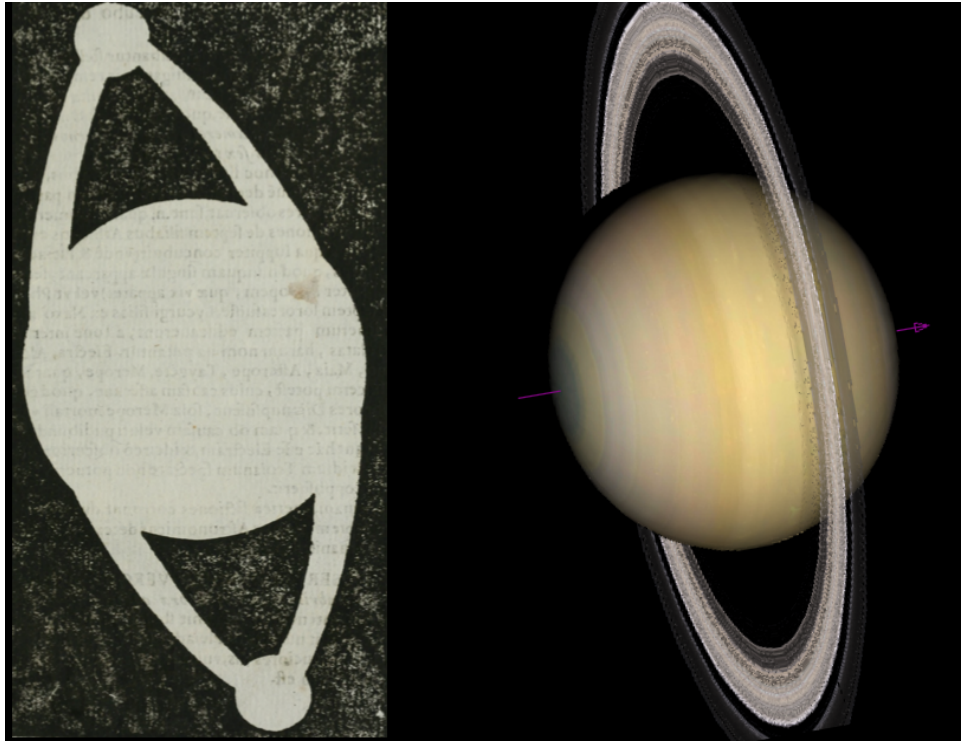

Fig. 11. Fontana observation of 12 Dec 1645 with the planet as seen from Naples in the same date. The axis of the simulation has been slightly adjusted by 10 degrees to match the drawing. The disk/body ratio and the tilt of the rings is rather well reproduced in Fontana's drawing.

### 7.2 The Pleiades observations

Fontana presented also the observation of stars in the field of the Pleiades. With one observation alone his telescope revealed 29 new stars. We recall that Galileo was able to see some 40 stars in the same field. However, no discussion or comparison is made here and Fontana only remarked that he believed the stars were countless.

## 8. Tractatus Octavus De Microscopio

In the opening of his book Fontana inserted a testimonial of the Jesuit Gerolamo Sersale who stated that he had used Fontana's microscope since 1625:

*I Jerome Sirsalis, Jusuit in the College of Naples, wish to bear witness to all that around the year 1625 in the house of this most illustrous gentleman, Francesco Fontana, the glory of his Neapolitan homeland, I saw a microscope, and after a short space of time, a telescope constructed by him with great skill from two convex lenses, so that such outstanding inventions, perceived by his divine genius, deserve to be reported.* (Translation Beaumont and Fay 2001).

In this section of the book Fontana described an instrument "*by which the smallest and virtually invisible things are so magnified that they can clearly and distinctly be examined*" made in 1618 (his fifth invention). The Lyncean Fabio Colonna informed Federico Cesi of the new invention by his friend Fontana on 17 July 1626 (cfr Freedberg 2002). Fontana did not pretend to be the first inventor of the microscope since "*the microscope could have been invented earlier elsewhere by someone else*".

Fontana then presented a detailed description of ten observations as an example of what he observed with the microscope. He describes a cheese mite, flea, an ant, a fly, several unknown animals, a spider, the sand, a human hair, material at the base of the window, and other things. As an example, the description of a cheese mite is below provided.

*The dust produced by cheese. This dust when placed under the microscope does not present the appearance of dust but of a remarkable living creature. It has eyebrows, lightly drawn as though painted with a brush, in like manner huge globes of eyes manifestly somewhat black, giving out a cheerful light. It is armed with little nails and claws, and seems to-be equipped with eyes. The entire appearance of its body too, in colour outstandingly exquisite, ennobles the tiny form of the animal, never before seen. To see it also -which cannot be done without marvelling - amounts to this: it crawls, feeds and definitely chews as well as moves itself; it seems equal in size to a human nail, its back is all rough and covered with scales, embellished with various star-like features, protected by thick and shaggy bristles, with such wondrous artfulness that you might have said that Nature, the creator of such a work, was born along with it, grew up with it, and even breathing with it, draws breath herself.* (Translation Beaumont and Fay 2001).

Federico Cesi and Francesco Stelluti in their Apiaria of 1625 made a first description of the anatomy of bees performed with microscopic observations. Apiaria was a gift to Pope Urban VIII and bore an attached engraving by Greuter, entitled the Melissographia, reproducing three bees as seen under the microscope. The arrangement of the bees was referred onto the trio of bees on the crest of Barberini family.
The word microscope was coined by Giovanni Faber in 1625 and the first printed microscopic illustrations were published five years later in the translation of the latin poet Aulus Persius Flaccus by Francesco Stelluti, *Persio tradotto in verso sciolto e dichiarato* (1630). On page 52 there is a reproduction of the 3 bees which closely remind the Greuter's Melissographia. We note note that at page 47 Stelluti writes that the bees' drawing *was observed and drawn by Francesco Fontana* [17], thus confirming that Fontana had a major role in the first microscopic observations of the bees. Colonna in his letters of 1626 to Cesi refers to Fontana as the friend of the Bee (Gabrieli, 1989).

The invention of the microscope is again quite unclear (cfr Zuidervaart 2011). Galileo made explicit mention of his microscope in the Saggiatore that was written in the period 1619-1622, though published in 1624, but he could have invented the miscroscope few years earlier. On 23 Sept 1624 Galileo sent an instrument that he named *occhialino*, to Federico Cesi with instructions on how to use it to see things closer. In the same year, the Kuffler brothers provided Cesi with a microscope made by Drebbel. Like the telescope also the microscope can have two optical configurations and it is quite possible that Fontana was the first to conceive a compound microscope made with convex lenses only.

## 9. Fontana and his telescopes in contemporary paintings

The paintings *Allegory of Sight* and *Allegory of Sight and Smell* by Jan Brueghel the Elder of 1617 and around 1618, respectively, show very sophisticated silver telescopes made with seven and eight draw tubes. It has been suggested that these telescopes are keplerians (Selvelli & Molaro 2009, Molaro & Selvelli 2011). This is deduced by the length of the telescopes which likely exceed two meters, and by the boxy shape of the eyepiece which is made to help the eye to be positioned precisely at the focus of the convex lens. As we have seen, in those years Fontana was the only one able to work two convex lenses in an accurate way to manufacture a Keplerian telescope. The telescopes in J. Brueghel The Elder's paintings belonged to the collection of scientific instruments of Albert VII Archduke of the Southern (or Austrian) Low Countries. Albert VII was brother of Emperor Rudolf II in Prague, protector of Kepler and Tycho, and brother of Maximilian III who, as we have seen, had a Keplerian telescope around 1616. All three Habsburg brothers were ruling the catholic Europe, to which the Kingdom of Spain and Naples was belonging. The viceroys in Naples were also fond of astronomy and of the military applications of the telescope and were in possession of the Fontana telescopes as documented in the letter of Colonna to Cesi of 30 Nov 1629. According to Lorenzo Crasso (1666) *Fontana made telescopes for all the courts and nobles around Europe which when obtained one of his telescopes conserved it together with the most precious things* [18]. Thus, it is quite possible that a preferential circulation of scientific instruments took place within the catholic countries, and that Fontana's instruments reached the far courts in northern Europe even before other places in Italy.

Brueghel's series was preceded by only a few years by another series of senses made by the Spanish painter Jusepe de Ribera who for his *The Sight* chose a telescope for the first time. We note here that the sitter in *The Sight* by Ribera holds a close resemblance with the self-portrait made by Fontana for his book. *The Sight* depicted by the young Ribera under the influence of Velasquez is shown in Fig 12, where a man is holding, though on the wrong side, a sophisticated telescope. Ribera's painting is not dated and according to Mancini (1956) is considered to have been executed during the end of the roman period of the painter sometime between 1613 and 1616. The canvas was commissioned by an unknown Spaniard and have been identified quite recently by Longhi (1966). Earlier the Allegory of Sight was attributed to Velasquez. Ribera arrived in Rome in 1611, perhaps already in 1608 and in May 1616 moved to Naples where in November he was married to the 16 year old daughter of the painter G.B. Azzolino. Such a quick acclimatization to Naples suggests that Ribera was familiar with the town and he could have possibly visited it before. It must be recalled that Pedro Téllez-Giron y Velasco, III Duke of Osuna was Spanish ambassador in Rome when Ribera was in Rome, became viceroy in Naples precisely in 1616 and was a patron of Ribera since the early days probably appointing him as a court painter. In my view that the Allegory of Sight could have been painted or finished in Naples is also suggested by the marine landscape depicted in the window which is similar to what could be seen from a window of a house in Naples.

The series of the five senses shows a caravaggesque naturalism with the figures represented with high contrast in the tradition of tenebrism painting. The two faces of the self-portrait by Fontana and the anonymous sitter in Ribera's painting are shown next to one another in Fig 13. Two other portraits of Fontana are reproduced in the Crasso (1666) and Terracina (1822) but are probably both derived from Fontana's self portrait. The shape of the head and the characteristic of the face and of the gaze are strikingly similar. One main difference between the two portraits lies in the hair. However, Fontana in 1646 represented himself as he looked in 1608, i.e. almost 40 years younger, and the simplest way to look younger is obtained by adding hair. Anyway the possible Fontana in the painting of Ribera should be few years older. Also the ears are different, but it must be considered that Fontana's self-portrait cannot be compared to those of one of the most talented painters of his times. Thus, though it is generally believed that Ribera took his models from everyday life it cannot be excluded that for the specific subject of the Allegory of Sight Ribera took inspiration by the figure of Fontana that in those years was already a renowned telescope maker. The difference between the expression of profound reflection of the *Allegory of Sight* with the drinker of the *Sense of Taste* has already been noted (Pérez Sanchez, 1992). The telescope decorated with gold is not something that can be associated to a street man since in those years it was very precious and a symbol of power. We admittedly prefer the possibility that he could be associated with the inventor of the *astronomical* telescope.

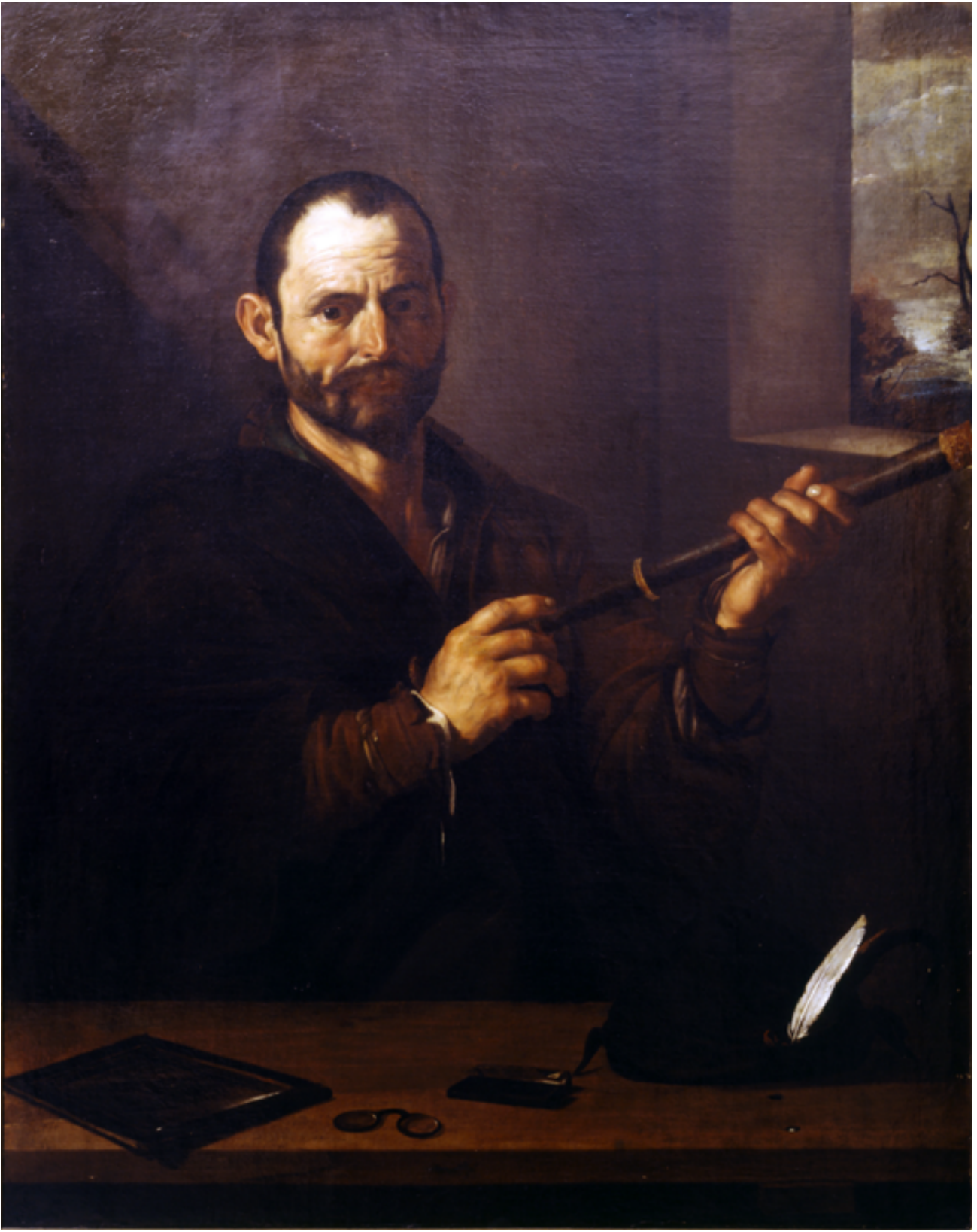

Fig. 12. Jusepe Ribera (around 1615) Museo Franz Mayer, Mexico City. Oil on canvas 114x89 cm. Courtesy of the Franz Mayer Museum.

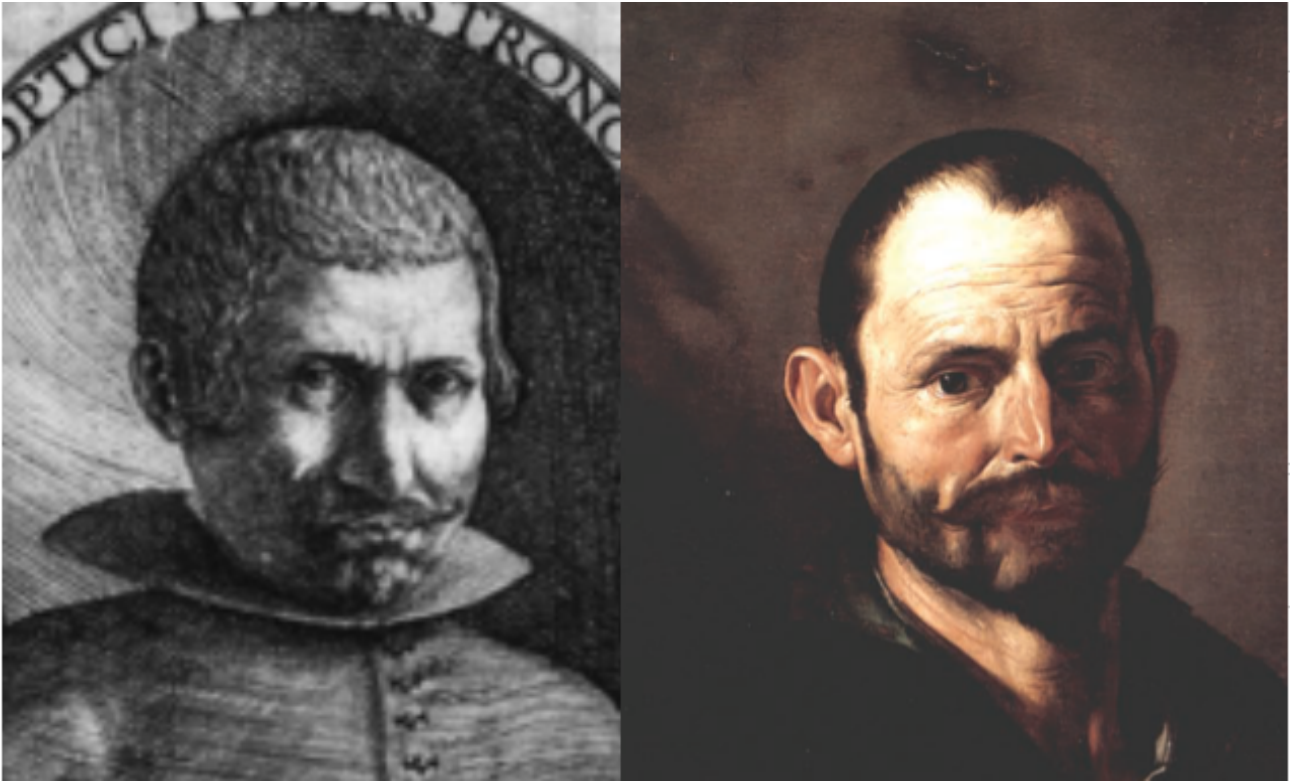

Fig. 13. Left: particular of the self-portrait made by Fontana in 1646 but referring to himself in 1608. Right: particular of the head of the sitter of The Sight by Ribera around 1615.

### Acknowledgements:

This work could not have been possible without the translation from Latin together with very useful annotation by Peter Fay, and Sally Beaumont who are warmly acknowledged though very sadly they cannot read this. I thank also Elisabetta Caffau for providing me access to the Fay and Beaumont's translation available at the library of the Observatoire de Paris; Simone Zaggia for his help in the use of Stargaze; Pierluigi Selvelli and Franz Daxecker for helpful discussions on the history of the telescope. Simonetta Fabrizio and Gabriella Schiulaz are warmly thanked for the English correction and Chiara Doz for help in the library search. Albert Van Helden is warmly acknowledged for his accurate refereeing.

### References

Arrighi, G. 1964 Gli occhiali di Francesco Fontana in un carteggio inedito di A. Santini, Physics, VI pp. 432-448

Beaumont S. and Fay, P., 2001, translation of the Novae coelestium, terrestriumq[ue] rerum observationes, et fortasse hactenus non vulgatae by Francesco Fontana. Private publication.

Castelli, B. 1637 Letter to Galileo of 10 Oct OG, XVII, 3572, p 191-192
Castelli, B. 1637 Letter to Galileo of 31 Oct OG, XVII, 3589, 208-209

Castelli, B. 1638 Letter to Galileo 3 July OG, XVII, 3753, p 350

Castelli, B. 1638   Letter to Galileo of 10 July OG, XVII, 3757, p 353

Castelli, B. 1638   Letter to Galileo of 17 July ON, XVII, 3759, p 355

Castelli, B. 1638   Letter to Santini of 18 Oct, cfr Arrighi,1964

Colangelo, F. 1834, Storia dei Filosofi e Matematici Napoletani, Trani, Napoli, VI pp 246-268

Colonna F. 1626 Letter to Cesi of 17 July, in Giuseppe Gabrieli 1996 Il carteggio linceo, Roma: Accademia Nazionale dei Lincei.

Colonna F. 1629 Letter to Cesi of 30 Nov, in Giuseppe Gabrieli 1996 Il carteggio linceo, Roma: Accademia Nazionale dei Lincei p. 1205.

Cozzolani G.G. 1638 letter to C.A. Manzini of 11 Sep. ON, XVII, 3783, pp 374-375

Crasso, L, 1666, Elogii d'hvomini letterati, Venice, Combi & La Nou, Tomo II, pp. 296-300

Daxecker, F. 2004 The Physicist and Astronomer Christopher Scheiner 2004, Publications at Innsbruck University, Vol. 246. Tyrolean State Museum Ferdinandeum (TLMF, Tiroler Landesmuseum Ferdinandeum, Innnsbruck, Museumstrasse): Initium et progressus Collegii Soc. Jesu Oenipontani. Litterae Annuae eiusdem Collegii 1561-1658. Dipauli 596/I, 1616, fol 41

Del Santo, P. 2009 On an unpublished letter of Francesco Fontana to the Grand-Duke of Tuscany Ferdinand II De Medici. Galilaeana, 6 pp 235,251

Favaro, A. Galileo e il telescopio di Francesco Fontana, in Atti e mem. Dell'Acc. Di scienze lett. E arti di Padova, n.s. XIX (1903), pp 61-71

Fontana, F. 1646 Novae coelestium, terrestriumq[ue] rerum observationes, et fortasse hactenus non vulgatae. A Francisco Fontana, specillis a se inventis, et summam perfectionem perductis, editae Neapolis, Gaffarum. ETH-Bibliothek Zürich, Rar 4215:1 http://www.e-rara.ch/doi/10.3931/e-rara-450 / Public Domain Mark.

Fontana, F. 1638 Letter to Vincenzo de Medici of 3 Jan

Freedberg, D. 2002 The Eye of the Lynx: Galileo, His Friends, and the Beginnings of Modern Natural History. University Chicago Press

Gabrieli G. 1989 Contributi alla storia della Accademia dei Lincei, Roma, 1989 347-371.

Galileo, G. 1638 Letter to Castelli of 25 July Edizione Nazionale Opere di GG, XVII, 3765,pp. 359

Galileo, G. 1638 Letter to Castelli Edizione Nazionale Opere di GG, XVII, pp. 360

Galileo, G. 1639 to anonymous of 15 Jan Edizione Nazionale Opere di GG, XVII, pp. 17-19

Huygens, C. 1888 Oeuvres complètes, I, La Haye, ad indicem

Longhi, R. 1966 I cinque sensi del Ribera Paragone N. 193 pp 74-78

King, H, C, 1955 The History of the Telescope, Dover books, Mineola, New York .

Kircher, A. Roma, Arch. della Pont. Univ. Gregoriana, Carteggio Kircheriano. XIII, f.33.

Kircher, A. 1646 De arte magna lucis et umbrae, Roma, pp.16. 831

Kragh, H. 2008 The Moon that Wasn't: The Saga of Venus' Spurious Satellite, Science Networks: Historical Studies 37

Malet, A. 2010, History of Science and Scholarship in the Netherlands, Vol 12, Amsterdam University Press, p 281.

Mancini, G. 1956 Considerazioni sulla Pittura (1614-21) Edited by Adriana Marucchi and Luigi Salerno, 2 vols, Rome

Micanzio, F. 1638, Letter to GG of 31 July , ON XVII 363

Molaro, P. 2013 On the earthshine depicted in the Galileo's watercolors of the Moon. 2013 Galileana Anno X, pp 73-84.

Molaro, P. Selvelli, P. 2011 A Telescope Inventor's Spyglass Possibly Reproduced in a Brueghel's Painting, ASPC 441, 13

Pérez Sa'nchez, A. 1992 Jusepe de Ribera The Metropolitan Museum of Art, NY pp35,165.

Polacco, G. 1644 Anticopernicus Catholicus, sev de terrae statione, et de solis motu, contra systema Copernicanum, Catholicae Assertiones, Venice, p 18

Renieri, V. 1639 5 March Letter to GG, ON XVII p. 308.

Rezzi, L. M. 1852 Sull'invenzione del microscopio Nuovi Lincei, V, pp.108 ss.

Riccioli, G. B. 1651 Almagestum novum, Bononiae, pp.203. 208. 485 ss-

Scheiner, C. 1616 Tractatus de Tubo Optico, unpublished (Munich, Bayrische Staatsbibliothek, MSS. Clm 1245,28-40)

Scheiner, C. 1630 Rosa Ursina apud Andream Phaeum Typographum Ducalem

Schyrleus, A. M. de Rheita, 1645 Oculos Enoch et Elliae Sive Radius Sydereomysticus, Antwerp: Hieronymus Verdussen.

Selvelli, P. Molaro, P. 2009 On the telescopes in the paintings of J. Brueghel the Elder. Proceedings of 400 years of Astronomical telescopes Conference at ESA/ESTEC in Noordwijk. Brandl, Bernhard R.; Stuik, Remko; Katgert-Merkeli, J.K. (Jet) (Eds.), (arXiv:0907.3745)

Stelluti, F. 1630 Persio tradotto in verso sciolto e dichiarato, Roma Giacomo Mascardi editor, p-47

Terracina G. 1822 Biografia dei Re e degli Uomini Illustri del regno di Napoli e Sicilia, a cura di Martuscelli, Vol 4, Napoli.

Torricelli, 1647 letter to Vincenzo Renieri of 25th May. Le opere dei discepoli di G. G. Carteggio 1642-58, a cura di P. Galluzzi - M. Torrini, I, Firenze 1975. P. 366

Van Helden 1976 The astronomical Telescope 1611-1650 Annali dell'Istituto e Museo di Storia della Scienza di Firenze, 1, no. 2, p. 13-36

Van Helden, A. 1977 The Development of Compound Eyepieces, 1640-1670, JHA, Vol. 8, p. 26-37

Van Helden, A. 1977  The Invention of the Telescope Transactions of the American Philosophical  Society, New Seres, Vol 67, No.4, pp 1-67

Van Helden et al. 2011 The Origins of the Telescope Amsterdam University KNAW Press.

Van de Vijver S.J., 1971 Lunar maps of the XVIIth Century, Vatican
Observatory Pubblications, Città del Vaticano, vol. 1, n. 1, p. 75.

Van de Vijver S.J., 1971  Original sources of some early lunar maps,  JHA, pp 86-97

Zuidervaart H. J. 2011 in  The Origins of the Telescope Amsterdam University ed Van Helden et al KNAW Press. p 9-46.

Winkler, M.G., Van Helden, A 1992 Isis, Vol. 83, No. 2, p. 195-217

Whitaker E.A 1999 Mapping and Naming the Moon a history of lunar cartography and nomenclature Cambridge university press

Footnotes

[1].  Dicendo sempre una stessa sentenza a chi li richiedea perchè non attendesse al patrocinio delle cause: che non bastandogli l'animo di trovar nel Foro la verità, sapea trovarla in quegli esercizi. (Crasso, 1666)

[2].  Questo [F.Fontana] per la relazione che ne ho, non è huomo di lettere, ma col continuo operare e fabricar cannocchiali si dice esser caduto in uno di tal singolarità, che per le cose del cielo è un miracolo (Micanzio 1638)

[3].  Havendone tra gli altri fabbricati per propria commodità  due di smisurata longhezza, adattolli su piè di legno nella sommità della casa, co' quali osservando continuamente le Pianete, ne formò quel Libro intitolato Novae caelestium terrestriumque rerum Observationes, da lui dato alla luce nel 1646 (Crasso).

[4] .  In these paragraphs we  maintain the original Latin titles   given by Fontana  in his book.

[5].  The theory of its construction is to be found in no earlier author than  in Book 17 of Johann Baptist Porta's Magic of Nature Chapter 10, printed 1589, which says this: "Concave lenses make distant objects clearly visible, convex lenses near objects [smaller?]…And that either Galileo put Porta's into practice, or he perfected it. (Translation Fay and Beaumont)

[6].  Porta holds the first realm;  German [i.e. Kepler!], you may have the second;
        your work, Galileo, gives you the third realm of the stars.
        But as far as the heavens are distant from the earth,
        you, Galileo, shine more brightly than the rest. (Translation Fay and Beaumont)

[7].  "mi ritrovo un occhiale di quelli di Napoli di gran perfezzione, e tale che non ho mai visto il meglio assolutamente….[]"     and    "mi vado intrattenendo con adorare l'occhiale meraviglioso veramente…centossessanta volte.. cosa mostruosissima"   Benedetto Castelli letter  to Galileo, 1637

[8].  "io mi ritrovo in mano un vetro di Napoli che serve per un cannone lungo quattordici palmi napoletani, [[] ingrandisce l'oggetto novanta volte"  letter of castelli to Galileo on 3 July 1638.

[9]. "Quanto al modo di lavorare le lenti napoletane; il vederle pulite esquisitamente non in tutto il disco, ma nella parte di mezzo, lasciando a tondo come una ciambella  non bene lustra, confonde il cervello a questi

artefici quà. Io ho pensato a qualche cosa di non triviale, ma non ardisco di aprir bocca, havendo altro per il capo." Galileo letter to Castelli of 20 July 1638.

[10]. "In the year 1590 the first tube was made and invented in Middelburg in Zeeland by Sacharias Janseen, and at that time the longest were 15 to 16 inches... The length of 15-16 inches was in use until the year 1618; then I and my father invented the long tubes which are used at night for seeing the stars and the Moon" Johannes Sachariassen answer to the Middelburg City Council investigation of 1655

[11]. "…If you fit two like [convex] lenses in a tube in the same way, and apply your eye to it in the proper way, you will see any terrestrial object whatever in an inverted position but with an incredible magnitude, clarity, and width". From *"Rosa Ursina sive Sol"* (1631) by Christoph Scheiner.

[12]. "opticum quodam instrumentum acquirerat admirandi usus, ita tamen ut imagines inversas redderet; quos cum Ser.mus [Serenissimus Maximilian III] rectas videre cuperet, nec que ratione id perficeret vel per alios reperiret" Manuscript of 1616, kept in the Tyrolean State Museum Ferdinandeum. Initium et progressus Collegii Soc. Jesu Oenipontani. Litterae Annuae eiusdem Collegii 1561-1658. Dipauli 596/I, 1616, fol 41.

[13]. "E' giunto a Genova un ritratto della luna, inviato qua dal P.D. Benedetto Castelli, con voce di un telescopio nuovo inventato da un tal Fontana di Napoli che mostra più esquisitamente le cose che non fanno i consueti." Letter by Renieri to Galileo on 5 March 1638

[14]. " in Napoli ci è una persona ingegnosa ma non have atteso a scienze. Se chiama Francesco Fontana,……Mando a V:P l'effige della Luna …osservata e disegnata dall'istesso Fontana. Di questi disegni ne sono andati in Roma al S. Cardinale Barberino, In Fiorenza al Gran Duca e forse ad altre persone ch'io non so."
"In Naples there is a person ingenious but has not studied science. His name is Francis Fountain, ...... I send to you the effigy of the Moon ... and observed also designed by Fontana. These designs have gone to Rome to S. Cardinal Barberino, Florence the Grand Duke and perhaps to other people that I do not know." Letter of Gloriosi to Santini on 13 March 1638.

[15]. Marte nel suo centro si scorge una prominenza come un velluto nero che termina in figura di cono e d'intorno ci stanno due cerchi o due fasce …e tutto ciò è mobile, atteso che non si mira sempre nell' istesso luogo..[] Letter from Cozzopani of 11 September 1638 to Carlo Antonio Manzini.

[16]. Quanto al pianeta Marte si è osservato che essendo al quadrato col Sole, ei non si vede perfettamente rotondo, ma alquanto sguanciato, simile alla Luna, quando ha 12 o 13 giorni, che dalla parte opposta a quella, che è tocca da i raggi solari, resta non illuminata, e per conseguenza non veduta: cosa che io dicevo dover aparire quando Marte fosse veduto superiore al Sole..[] Galileo's letter of 15 January 1639 to unknown.

[17]. "il tutto ancora esquisitamente osservato e disegnato il Signor Francesco Fontana: onde feci qui in Roma intagliare in rame tre Api rappresentati l'Arme di Nostro Signore Papa URBANO VIII" Francesco Stelluti, *Persio tradotto in verso sciolto e dichiarato* (1630) page 47.

[18]. "per lo chè da più Sovrani Principi priegato de' suoi Telescopii, e con longhezza di tempo, e favor grande, ricevuti, venivan subito riposti tra le cose più pregiate nelle lor Galerie (Crasso)